\documentstyle[preprint,eqsecnum,aps]{revtex}
\tightenlines
 5
\begin{document}
\draft
\title{Pseudoscalar meson 
photoproduction: \\
from known to undiscovered resonances}   
\author{Bijan Saghai}
\address{
Service de Physique Nucl\'{e}aire, CEA/DSM/DAPNIA, 
Centre d'Etudes de Saclay, \\
F-91191 Gif--sur--Yvette, France
}
\author{ Frank Tabakin~\cite{byline} }
\address{
Department of Physics \& Astronomy,
University of Pittsburgh,
Pittsburgh, PA  15260
}
\date{\today}
\maketitle   
\begin{abstract}
The role of  dynamics in
 spin observables for pseudoscalar meson photoproduction 
is investigated using a density matrix approach in a multipole 
truncated framework. Extraction of novel rules for   
$\gamma p \rightarrow \pi^+ n,~ K^+ \Lambda$ and 
$\eta p$ reactions  based on resonance dominance,
and  on other broad and reasonable dynamical assumptions, are discussed. 
Observables that are particularly sensitive to
missing nucleonic resonances predicted by quark-based approaches,
are singled out. 
 
\end{abstract}
\narrowtext
\section{ INTRODUCTION}

Determination of the dynamics
underlying  pseudoscalar 
meson photoproduction has been a major 
challenge in hadronic physics for several decades. 
This challenge persists because: 
{\it i)} data remain
scarce and of rather poor quality (except perhaps for the pion
production case); and {\it ii)} the most advanced approaches, 
based on effective Lagrangian formalisms, embody entities not 
calculable {\it via} a fundamental theory, and hence require 
free parameters.

Intensive experimental effort at the Continuous 
Electron Beam Accelerator Facility (CEBAF),  the 
 Electron Stretcher Accelerator (ELSA),
the European Synchrotron Radiation Facility (ESRF), 
the Laser Electron Gamma Source (LEGS) and
at the Mainz Microtron (MAMI) are, or will soon be, providing copious
and accurate data. One major anticipated advance is the  
measurement of single and double polarization observables.
Simultaneously, phenomenological theories are becoming 
more sophisticated. Nevertheless, for  
kaons, and to a lesser extent for $\eta$ and $\pi$    
photoproduction, a unique determination of the underlying dynamics is 
not anticipated because of possible contributions from a 
rather large number of  resonances to the reaction
mechanism.

The present work is motivated by an effort to ameliorate
 this awkward situation.  We  offer a potentially useful  
  link  between forthcoming polarization data and 
phenomenological analysis.
In generating this link,  we start from the model independent rules
of Ref.~\cite{FTS}.  In applying those rules 
  to specific reactions,  we invoke some
  broad and reasonable dynamical assumptions.
These assumptions are: (1)  the multipole amplitudes can be truncated, 
based on the centrifugal barrier; (2) multipole amplitudes are 
resonance dominated; and (3) the background and non-resonant
contributions are small and structureless.  Some of these are bold 
assumptions,  but they do allow us to generate
 guidelines for resonance searching,  prior to a full-fledged dynamical 
calculation.
Thus, the purpose of this paper is to  extend our earlier 
study~\cite{FTS} of {\it nodal structure}  to isolate 
specific dynamical features.

Each pseudoscalar meson photoproduction case (
$\gamma N \rightarrow \pi N,~ $KY$,~\eta N$), is known to have 
different characteristics.
Pion photoproduction is the
best understood channel. It has the advantage of being
dominated by {\it only} one nucleonic resonance 
$\Delta_{33}.$  We show later that 
 our analysis of spin observables agrees with the results of the best 
available phenomenological formalisms and sheds some light on further
 developments. 

Among the three pseudoscalar meson production processes, the reaction 
mechanism for strangeness production is the most complicated
and hence understood the least. This reaction has been discussed
in detail in a previous paper~\cite{ST}.  For this reaction, 
 we now provide more information by focusing on
 very recent polarization data and show how our 
nodal trajectory analysis deepens understanding of
 recent phenomenological models.

 Finally,
we study the $\eta$ production case. Using recent
experimental and theoretical results, we show here how 
the $\eta$ production process might be used to search for missing,  
or undiscovered,  resonances.~\footnote{Missing, or undiscovered, 
resonances have been investigated by several authors. For 
 illustration, we refer only to   recent papers by Capstick 
and Roberts~\cite{cap94,cap93}, which contains references to 
other relevant works.}   These resonances are   
predicted~\cite{cap94} by quark-based models~\cite{cap94,cap93,kon80} 
to couple only weakly,  if at all, to $\pi N$ systems,  but 
significantly to the $\eta N$ channel,  which enhances interest 
in $\eta$ production.  

In  Section II, the general structure of the cross section and 
all fifteen single and double polarization observables is presented. 
The notion of nodal trajectories is illustrated and applied to 
specific cases in Section III. Our conclusions
are presented in  Section IV.


\section{ Spin observables in a multipole truncated basis and
dynamical rules}

The general rules for the sixteen observables, derived from a 
density matrix approach, are described
 in detail in Ref.~\cite{FTS} (FTS).
From that work, we recall that 
the Legendre classes of the sixteen observables,
 which are labeled by ${\cal L}_0$, ${\cal L}_{1 a}$,
${\cal L}_{1 b}$, and ${\cal L}_2$, are: 
$${\cal L}_0 ({\cal I}; {E}; {C}_{z'}; {L}_{z'}),~   
{\cal L}_{1 a} ({P}; {H}; {C}_{x'}; {L}_{x'}),~   
{\cal L}_{1 b} ({T}; {F}; {O}_{x'}; {T}_{z'}),~
{\cal L}_2 ({\Sigma}; {G}; {O}_{z'}; {T}_{x'}).$$

In the above list, as explained in Table I,
the first entry in each class is the cross-section
or a single
polarization observable
 $( {\cal I}, {P}, {T}, {\Sigma});$
  the others are all double polarization
observables,  which appear ordered as Beam-Target
 $( {E}, {H}, {F}, {G}),$
Beam-Recoil  
 $( {C}_{z'}, {C}_{x'}, {O}_{x'}, {O}_{z'});$
with the last entry in each class being the Target-Recoil
observables
 $( {L}_{z'}, {L}_{x'}, {T}_{z'}, {T}_{x'}).$
The polarization asymmetries range from  $-1$ to $+1.$ 
 The angular dependence of the 
 above observables  are determined  by expressing the four  
 helicity amplitudes $H_i (\theta)\  (i=1 \cdots 4)$ in terms of   
 Wigner rotation functions, with $\theta$ denoting the produced meson's
center-of-mass angle. 
It is then simple to deduce that each  
${\cal L}_M$ class
observable can be expanded in a series of associated Legendre   
functions $P_{L M} (\cos \theta) .$

Rules concerning spin observables were discussed by FTS, based on  
the possible truncation of helicity or multipole amplitudes. 
The advantage of expanding the meson photoproduction amplitudes
 into multipoles
$E^{\pm}_{\ell}, M^{\pm}_{\ell}$ is that the orbital angular moment, 
$\ell,$ of the final {\it meson-baryon} state can be used to  reduce the
number of amplitudes,  based on the existence of a centrifugal barrier. 
Of course, this  truncation does not include the possibility of 
dynamical effects, which could magnify selected orbital states. 
For example, a resonance could emphasize a particular partial wave  or 
competing effects could attenuate selected waves. However, it
is just the deviation from ordinary centrifugal-dominated behavior
of spin observables  and the dominant role of
baryonic resonances that allow spin observables 
to  serve as  excellent indicators of special dynamical effects.

Spin observables organized
by Legendre class and expressed as
profile functions\footnote{
Profile functions~\cite{FTS} are the product of the
spin observable times the cross-section function
${\cal I},$ with the cross-section
given by $\sigma(\theta) =(q/p){\cal I},$ where
$p$ and $q$ denote the initial and final state c.m momenta. 
Profile functions are proportional to bilinear products of amplitudes. } 
are expanded in the following forms.
For members of the Legendre class ${\cal L}_0,$ the form is:
$$ {\cal O} \equiv \sum_{L \geq 0} A_L P_L(\cos \theta)
 \equiv 
\sum_{m \geq 0}^n a_m  \cos^m\theta.$$ 
For members of the Legendre class ${\cal L}_{1a} $ or 
${\cal L}_{1b}$, the form is:
$$ {\cal O} \equiv \sum_{L \geq 1}  A'_L P_{L  1}(\cos \theta)
 \equiv  \sin\theta\  
\sum_{m \geq 0}^n a'_m  \cos^m\theta.$$    
For members of the Legendre class ${\cal L}_2 $    
the form is:
$$ {\cal O} \equiv \sum_{L \geq 2} A''_L P_{L  2}(\cos \theta)
 \equiv  \sin^2\theta\  
\sum_{m \geq 0}^n a''_m  \cos^m\theta.$$  The coefficients
$a_m, a'_m, a''_m$ can be expressed in terms of the basic
multipole amplitudes.  The manner in which a specific
multipole contributes to these coefficients,  and the
possibility that the associated polynomial can have
nodes,  are the major features that we exploit in this paper
to deduce definitive manifestations of underlying
hadron dynamics.  For example, the condition for nodes, $\theta_0,$
in a spin observable, aside from  endpoint
$(0^\circ\ \&\ 180^\circ)$ zeroes, is 
 $\sum_{m \geq 0} a_m  \cos^m\theta_0 = 0. $
 
For example, under the assumption that the $a_{m \geq 3}$ coefficients
can be neglected, one needs to consider the quadratic equation
$a_2 x^2 + a_1 x + a_0 =0 $ ( $x=\cos\theta),$
 which has two solutions:
 $x_{1,2} = [-a_1 \pm (a_1^2 - 4 a_0 a_2)^{1/2}]/{2 a_2}.$
 Nodes
occur if a root is real and less than 1 in magnitude.
 To get real solutions, 
we need  $a_1^2 \geq 4 a_0 a_2.$  For
$a_1^2 = 4 a_0 a_2,$  we can get two equal solutions
$x_1=x_2 =  -a_1/ 2 a_2;$  if these solutions are less than 1 
in magnitude, then the observable has a non-sign-changing zero (NSC),  
not a sign-changing (SC) node. That locates the bifurcation
point, which is the energy at which double nodes first set in.
  One can also generate conditions on
the derivative of the profile function with respect to $\theta,$
which can be used to test if $m\geq 3$ coefficients can be neglected.
Using such features, knowledge of nodes in a spin observable can 
 provide definitive information
about the $ a_0, a_1, a_2 \cdots$ coefficients and thus about 
underlying multipole amplitudes and resonances. 
Of course, by fitting data directly over
a range of energies,
one can extract even more information from these coefficients.

To constrain  dynamics and  
the basic multipole amplitudes,  it
is useful to express the coefficients $a_m$ in terms of the
 electric and magnetic multipoles.  The basic idea here is that for 
each photoproduced meson there is a family of dominant resonances.  
Those resonances feed into the multipole amplitudes of the same quantum 
numbers,  which, in turn, determine the polynomial coefficients, $a_m.$  
Once these are known,  the general energy and angular dependence of
all spin observables, along with associated nodes, can be specified.  
Thus, each meson has spin observables characterized by its driving 
resonances.

To illustrate the angular dependence and the energy evolution
of nodes, the spin observable $E,$ a typical ${\cal L}_0$ 
Legendre class observable, is shown in Fig.~1. At the
lowest incident energy,  $E$ takes only positive values;
with increasing energy, it assumes a zero value at one angle
(a non-sign changing zero, NSC).  As the energy increase that single
 zero  bifurcates into two nodes (sign changing nodes, SC).
The projection of these nodes into the {\it node position-energy}
plane constitutes the {\it nodal trajectory}.

 With this 3D-plot in mind,
 we would like to stress two features used extensively
 in this paper. {\it The polynomial behavior of 
the angular 
distribution of the observable and/or its nodal structure depends on 
 the  incident photon energy.  Moreover, such dependence may 
(and often does) vary from one observable to the other in a manner
characterized by the underlying resonances.}  This remark can be 
applied to every pseudoscalar meson photoproduction reaction.


\section{ Nodal trajectory analysis}

Expressions relating the
coefficients $a_m, a'_m$  and $ a''_m $ 
to electric and magnetic multipole 
   amplitudes (truncated at $\ell \leq 2$)  were
obtained~\cite{GST} 
for all sixteen observables (${\cal O}$)  
using Mathematica. These observables are organized as 
${\cal O}_M =  \sin^M\theta \sum_{m \geq 0}^n a_{m}x^m$, with 
$x \equiv \cos(\theta),$ where $\theta$ is the meson-baryon final-state
angle in the c.m. system and the 
label $M = 0, 1,$ or $2,$
for  Legendre class 0, (1a, 1b), or 2, respectively.  In 
Appendix~A, a sample result is presented for the target polarization 
profile function.  The relevant $a_0 \cdots a_5$ coefficients are given 
as imaginary parts of bilinear products of multipole amplitudes. 
By examining the structure of this particular result, we can understand 
the general form of all spin observables, as displayed in Table~II. 

To understand the notation used in Table~II,  consider the
$a_0$ for the target polarization profile
 given in Appendix~A.  Its first term involves the 
S-wave multipole $E^0_+ ,$ which has a total angular momentum
of $J=1/2,$ and can be designated as an $S_{2I,1}\equiv S$ amplitude, 
using the usual convention $L_{2I,2J}.$  For 
convenience, we  present the case of general isospin $I.$  The
$E^-_1$ and $M^-_1$ multipoles are P-wave amplitudes
 with $J=1-1/2=1/2$
and thus are designated as $P_{2I,1}\equiv P$ amplitudes.  Similarly,
$E^+_1$ and $M^+_1$ are P-wave amplitudes with $J=1+1/2=3/2$
and thus are designated as $P_{2I,3}\equiv P'$ amplitudes.  
 For D-waves, we have $E^-_2$ and $M^-_2$  amplitudes  
with $J=2-1/2=3/2,$ designated as $D_{2I,3}\equiv D$
 and $E^+_2$ and $M^+_2$  amplitudes 
with $J=2+1/2=5/2,$ designated as $D_{2I,5}\equiv D'$ amplitudes. 
 
The first term of $a_0$ in Appendix A involves
interference between S- and P-waves of $J=3/2$. To highlight that feature, 
we abbreviate that term as:  ``$S P',$" where the prime
indicates again that the P-waves are of the $P_{2I,3}$ type.
  The remaining  terms  in
$a_0$ for $T$ involves   P- and D- wave interference; they include
$P_{2I, 1}$ ($M_1^-$) interfering with $D_{2I,3}$ and $D_{2I,5}$ terms
and also $P_{2I, 3}$ ($E_1^+\, \&\, M_1^+$)
interfering with $D_{2I,3}~(E_2^-\, \&\,M_2^-)$ and 
$D_{2I,5}~(E_2^+\, \&\, M_2^+)$ terms. 
These terms are  abbreviated as ``$P_{2I,i} D_{2I,j}$'' where 
$i$ takes on the P-wave $2J$ values of 1 and 3, 
and   $j$ the D-wave $2J$ values of 3 and 5.

For the  $a_1$ term of the target polarization, there are 
``$S D_{j}\equiv S D \oplus SD'$" terms, e.g. $S~\&~D$ interference 
involving $J=3/2$ and $J=5/2$ D-wave multipoles. In addition,
 interference  between   $J=1/2(P)$
and  $J=3/2(P'),$  is designated as a ``$P P'$" contribution.
Terms that involve $J=3/2(E_2^-\, \&\,M_2^-)$ interference with 
 $J=5/2(E_2^+\, \&\, M_2^+)$  D-waves, are expressed in Table~I 
as ``$D D'.$"  Terms in $a_2$  involving P-waves of the same 
angular momentum,  albeit of different electric or magnetic
multipole character, such as $E^+_1 M^{+*}_1,$ are denoted
by a single letter ``$P'$."  For corresponding D-wave terms,
$E^\pm_2 M^{\pm*}_2,$ for $ J=2 \pm 1/2$ cases, we enter a single letter: 
$D_{2I,3} \, \&\, D_{2I,5}\equiv D_{2I,j};$ where $j=3,5$ in Table~II.

At this stage,  we hope the compact notation used in Table~II
to describe the general structure of the coefficients $a_m$
is clear, since it is essential for the rest of our paper. 
 Note that there is an odd/even parity rule for the
$a_m$ terms in Table~II. For example, the $a_0$ for the target polarization
entry involves $\ell=0\ \&\  1$ plus $\ell=1\  \&\  2$ interference--which
are odd terms.  The next term, $a_2,$ is an even term which involves
$(\ell=1) \times (\ell=1)(P')$ and $(\ell=2) \times (\ell=2)(D_{2I,j})$ 
terms, plus three other manifestly even interference terms. That pattern, 
which appears throughout the table,  is clearly a reflection
of the underlying tensorial and parity character of each
spin observable.  Another important feature is the location
of the  $E^+_0$ multipole amplitude,  which is indicated
by the boxed terms in Table~II; interference and  also
magnitude, $``S",$  type terms appear in these boxes. 
As we will see in Sec.~III.A.3, this S-wave
amplitude if quite large, will make the $a_m$ in which it appears
dominant.  Terms that involve interference between this sizable, S-wave 
amplitude and particular P- and D- waves,
 have amplified values of the $a_m$ in which that S-wave occurs.  
If the S-wave interferes with a resonant P- or D-wave amplitude,
then great magnification of that term can occur;  which can cause
dramatic changes in nodal and polynomial structure.  
It is such a mechanism that we seek
to isolate and  use to magnify the role of as yet unseen resonances!

By identifying
different resonances, according to their angular momentum and spin,
with the relevant multipoles, the expressions for the 
observables summarized in Table~II can be used to anticipate the
role of resonances on spin observables. There are 
basically two types of terms in Table~II (see Appendix A):
 those coming from a single
resonance ($\propto |E^{\pm}_{\ell}|^2$, $|M^{\pm}_{\ell}|^2$, 
$E^{\pm}_{\ell} \cdot M^{\pm}_{\ell}$) and those arising
from interference terms between two resonances.  Here we begin to
identify amplitudes with resonances;  indeed, our key point is that
by {\it assuming that amplitudes are dominated by resonances},
we can anticipate the angle and energy dependence of spin observables
and their sensitivity to particular resonances.

One needs to be careful about treating the isospin.  Although Table~II 
refers only to a fixed isospin $I,$  the Table generalizes to the
$I=1/2 \  \&\  3/2$ case, as occurs for pions.  The interpretation of 
the P- and D-wave interference term ``$PD,$" maps to sums over 
isospin; namely,
$$PD \rightarrow  \sum_{I,I'}\ P(I) \cdot D( I')  ;$$
  whereas, the diagonal-type term ``$D$"
becomes~\footnote{Isospin factors and relevant phases have been 
incorporated into the amplitudes.}
 $$|D|^2 \rightarrow \sum_{I, I'}\ |D(I) +D(I')|^2 . $$

We can now apply the general rules for the structure of the
spin observables to different pseudoscalar mesons.

\subsection{ Dynamical Rules}

In this Section, we give examples of  
the angular distribution of the polarization observables in 
$\gamma p \rightarrow  \pi^+ n, \eta p$, and $K^+ \Lambda$ 
processes, in order to gain insight into how 
the above general rules help to reveal the basic dynamics.

\subsubsection{ Pion}

Pion photoproduction is by far the most investigated~\cite{VPI} of 
these reactions. Despite this attention, complete angular distribution 
data for polarization observables remain scarce. Experiments recently 
completed  at Bonn~\cite{BONN,buc94} and Brookhaven~\cite{BROOK} will 
soon greatly enlarge the data base. Here, we investigate the 
preliminary results from the PHOENICS Collaboration~\cite{BONN}.

The target polarization asymmetry, $T,$ at 
$E_{\gamma}^{lab}=220$ MeV and $650$ MeV for the reaction
$\gamma \vec{p} \rightarrow  \pi^+ n$ are
shown in Fig.~2. The profile function for this 
${\cal L}_{1b}$ Legendre class single spin observable 
can be expressed in the
form  $\sin \theta \times (a_0 + a_1 x + a_2 x^2 + \cdots)$.\
 The polynomial coefficients $a_m$ were adjusted to fit these data.
The data at both energies, Fig.~2, are quite well
 reproduced by a second order polynomial, for the higher 
energy results the need for a third order polynomial is unclear.
What can we learn from the fact that the data require a second 
order polynomial? 

We now use Table~II for the spin observable $T.$  
Based on the most recent phenomenological calculations, 
 we assume zero $D'\equiv D_{15}$ contribution.  In that case.
 the entry for $T$
 in Table~II shows the following coefficient
structure:
\begin{eqnarray}
 a_0~&=&~SP'~ \oplus~ PD~ \oplus~ P'D, \nonumber \\
 a_1~&=&~P' \oplus D \oplus~ SD \oplus PP', \nonumber \\ 
 a_2~&=&~P'D, \nonumber \\
a_3~&=&~0 .
\label{coefT}  
\end{eqnarray}
From the above, one sees that neglecting $D'$ resonances leads  
 to a second order polynomial. Moreover, 
to generate a nonzero $a_2$ the pion-nucleon system must have
significant, and we assume resonant, $P'D$ contributions.
Note the dominant $\Delta_{33}$ isobar is a $I=3/2,$ $P'$ state.
 The presence
of the $\Delta_{33}$  {\it and}  spin-3/2 ($l=2$) resonances
($P'$ and $D$, respectively)
are necessary to get $a_2 \not= 0.$  Thus, evidence for an $a_2$ 
polynomial,  under the assumption of zero $D'$
terms, can shed light on the role of a $D$ contribution.

Fitting the lower energy $220$ MeV data with a second order polynomial,
we find that 
  $a_2~\simeq~2 \cdot a_1 ;$ with 
$|a_2|$ slightly larger than $|a_0|$. Given that $a_1$ is the only 
coefficient containing a pure contribution from the dominant
$\Delta _{33}$ resonance (the single $P'$ term), the smallness of this 
coefficient implies that the other terms in $a_1$  interfere destructively 
with the $P'$ term. Also the extra terms in $a_0$ compared
to $a_2$, $SP'~ \oplus~ PD$, are slightly destructive, since we find that
$|a_2|$ is slightly larger than $|a_0|.$   Recall that to get a real
node in a second order polynomial,  we need $a^2_1\geq 4a_0a_2,$ which is 
not satisfied here; hence, this observable has its 
nodeless behavior despite the $\Delta_{33}$ resonance. 

 This absence of
nodes at $220$ MeV also implies that resonances other than the 
$\Delta _{33}$ are required by the data, as is already known from 
existing models (see, for example Ref.~\cite{gar93}).
 In particular, the $220$ MeV data yield values
of the polynomial coefficients which, from the above structure, require 
contributions from spin 1/2 ($S$ and $P$) and spin 3/2 ($P'$ and $D$)
nucleonic resonances. Again, we assume that the multipole
amplitudes are resonance dominated, although it is possible that a
background can play a significant role and should be included in
a fully dynamical analysis.

At the higher energy $650$ MeV, the absolute values of the coefficient 
$a_0$ for  both the $n=2$ and $n=3$ polynomial fits are small. 
   Without an $a_0$ term,
the observable $T$ has a $\sin\theta \times \cos\theta$ structure;  hence, 
a node appears in $T$ near $\approx 90^\circ$ for small $a_0.$  To obtain 
that small value of $a_0$ and the $\approx 90^\circ$ node, the terms in
the expression given above must interfere destructively, e.g.
$~SP'~ \oplus~ PD~ \oplus~ P'D \approx 0.$

Note that using Table~II, if the data requires $n=3$ terms,  then  
besides the spin 1/2 and spin 3/2 resonances, the
reaction mechanism  would acquire
contributions from a spin-5/2 resonance ($D'$).
Such a resonance, at higher energies ($\geq 800$ MeV), has been 
suggested by Garcilazo and Moya de Guerra~\cite{gar93} in their 
extensive study of pion photoproduction using an 
effective-Lagrangian-based model which includes   
{\it s}-channel, spin-1/2 and spin-3/2 resonances ($S,~P,~P',~\&~D$).  

Although,  we can learn from the above how to
analyze the general structure of $T$ for its resonance dependence, 
we also see from Table~II that the target asymmetry T is not
 the best observable for investigating 
the role of spin-5/2 resonances ($D'$). 
From Table~II,  the Beam-Recoil  
$( {C}_{z'}, {C}_{x'}, {O}_{x'}, {O}_{z'});$ 
and Target-Recoil  $( {L}_{z'}, {L}_{x'}, {T}_{z'}, {T}_{x'})$
 double polarization observables offer much cleaner
cases for that purpose. These observables can be classified in three
groups according  where a pure magnitude $D'$ term occurs.
For ${\bf C_{z'}} \& {\bf L_{z'}}, ``D"$ occurs in the $n=5$
term; for ${\bf C_{x'}} , {\bf L_{x'}},  
 {\bf O_{x'}}, \& {\bf T_{z'}}, ``D"$ occurs in the $n=4$
term;  while  ${\bf O_{z'}} \& {\bf T_{x'}}$ have the lowest
occurrence of a $``D"$--in their $n=3$ terms. 
The common feature to all of these double spin observables is not only  
that the highest power coefficient ($a_{n_{max}}$) is a {\it pure} 
$D'$ state, but also that 
 the $a_{n_{max}-1}$  coefficient depends {\it only} on the 
$P'D'$ interference terms. Given the dominant role played by the 
$\Delta _{33}$-resonance ($P'$), the effect of the $D'$ resonance 
is hence magnified in all of these observables,  with evidence for
 $a_2, a_3$ terms in ${\bf O_{z'}} \& {\bf T_{x'}}$  offering the
 best choice among these double spin observables.

 However, the most promising observables in looking
for the effects of spin-5/2  resonances are, according to Table II,
reached using a linearly polarized beam, e.g. 
the single polarization ${\Sigma}$ 
and double beam-target $G$ asymmetry.  
That conclusion is based on the fact that $D'$ enters into
 the lowest order polynomial terms for these observables.

Having shown how  Table~II, provides
 a guide for resonance searching in pion photoproduction,
 we now turn to another
example.

\subsubsection{ Kaon}

Due to a large number of resonances and 
{\it t-}channel exchanges, 
 the reaction mechanism  for associated strangeness photoproduction
is much more complicated~\cite{AS,WJC,SL,li} than for $\pi$ and 
$\eta$ photoproduction.
However, in $K^+ \Lambda$ (and $\eta$) channels only isospin
$I=1/2$ resonances can intervene,  which is at least one simplification
compared to the pion case.

The only published angular distribution data for polarization observables
are the hyperon-recoil ($P$) asymmetry recently measured at 
ELSA~\cite{ELSA}.
 In Figure~3 their results for the 
$\gamma p \rightarrow K^+ \vec{\Lambda}$ channel at 1.2 GeV are shown.
 In Fig.~3(a),  the results of our polynomial fit using the form: 
$P = \sin\theta  \sum_{m=0}^n a_{m}x^m$, with 
$x \equiv \cos(\theta_{cm}^K)$ are depicted for four polynomial orders 
($n= 1, 2, 3,$ and $4$). The end points are required to be  zero, by 
virtue of the helicity amplitude structure of this observable~\cite{FTS}.
From Fig.~3(a), we infer that although an $n=2$ polynomial gives an 
acceptable description of the data, the use of an $n=3$ 
polynomial decreases the $\chi^2$ by roughly a factor of 4, while
 there is no significant need for $n=4$ terms.
The structure of the $P$-asymmetry (see Table~II), shows that evidence 
for an $n=3$ polynomial implies the presence of spin-5/2 nucleonic 
resonance(s) ($D'$) in the underlying dynamics.

To confront this finding with our present knowledge of the relevant 
reaction mechanism, we show in Fig.~3(b) the {\it predictions} of 
three recent phenomenological approaches~\cite{AS,WJC,SL} based on 
isobaric formalisms. These
effective-Lagrangian-based models contain {\it s-}, {\it u-}, and
{\it t-}channel exchanges. In a previous paper~\cite{ST},  we 
investigated the implications of  these exchange 
channels on our nodal
trajectory analysis. Here, we will concentrate on the {\it s-}channel
nucleonic resonances. The {\it s-}channel content of the three 
models discussed here, 
can be summarized as follows. The two first models by Adelseck-Saghai 
($AS$)~\cite{AS},
and Williams, Ji, Cotanch ($WJC$)~\cite{WJC}, include only spin-1/2 
resonances. 
Namely, $AS$ : $ [P_{11}(1440)] \subset [P]$; 
$WJC$: $[S_{11}(1650),~ P_{11}(1710)] \subset [SP]$.
While the most recent model from the Saclay-Lyon Group ($SL$), 
by David {\it et al.}~\cite{SL},
contains  spin-1/2, spin-3/2 and spin-5/2 resonances: 
$ [ P_{11}(1440),~ P_{13}(1720),~ D_{15}(1680) ] \subset [ PP'D'].$  
From Fig.~3(b),  and the
$\chi ^2$ per point values 
($AS:2.5,~ WJC:1.9,~ SL:1.5 $), we conclude that 
the {\it genuine} spin-5/2 resonance in the $SL$
model is producing the anticipated $a_3$ effect discussed
 in the previous paragraph.

Note that the $AS$ and $WJC$ models reproduced the  
data with reasonable accuracy.  The $AS$ model {\it predicted}  the 
existing (old)
P-asymmetry data especially well,
 and the WJC model reproduces correctly all data included 
in their fitted data base.
As discussed in a previous paper~\cite{ST}, 
the  higher spin resonances missing 
in the $AS$ and $WJC$ models are mimicked by the $t-$channel exchanges, 
in line with the duality hypothesis. Nevertheless,
as we anticipated~\cite{ST},  the $P$-asymmetry is basically a 
resonance driven entity. This resonance dominance is confirmed by the 
results shown in Fig.~3.

For further {\it illustration} of the role of the polynomial 
coefficients, we mention that the  numerical values of the 
coefficients for the $n=3$ polynomial fit to the 
$\Lambda$-polarization asymmetry 
in Fig.~3(a) satisfy the following relations, at the level of 
a few percent: $|a_0| \simeq |a_2|$ and $|a_1| \simeq |a_3|.$ 
From Table~II, we see that:

$$\delta _{02} \equiv |a_0|-|a_2|\propto SP_{1i}~\oplus~ PD
\equiv SP ~\oplus~ SP' ~\oplus~ PD,$$ 

\noindent and 

$$\delta _{13} \equiv |a_1|-|a_3|\propto SD_{1j}~ \oplus~PP'
\equiv SD ~ \oplus~ SD' ~ \oplus~ PP' .$$
 
Let us now examine how our fit using the above $n=3$ polynomial
structure, which implies 
$\delta _{02} \simeq 0$ {\it and} $\delta _{13} \simeq 0,$ can arise. 
 
In a rather {\it complicated} reaction mechanism which includes 
$S_{11}$, $P_{11}$, $P_{13}$, $D_{13}$, and $D_{15}$ nucleonic 
resonances, the above relations 
($\delta _{02} \simeq \delta _{13} \simeq 0$) 
can be satisfied in one of the
two following ways:

{\it i) Strong interference effects} : 
highly destructive interference occurs among the $SP$, $SP'$, and $PD$
 terms, {\it and} also among  $SD$, $SD'$, and $PP' $ terms.  

{\it ii) Weaker interference effects}: 
if the contributions from $P_{11}$ resonance(s) are negligible,
then $\delta _{02} \simeq 0$ {\it and} $\delta _{13} \propto SD_{1j}$. 
In this case, either small $SD$ {\it and} $SD'$ or destructive interference
between these two terms will ensure $\delta _{13} \simeq 0$.
 
Actually, the $SL$ model (obtained within the most comprehensive 
phenomenological approach), is very close to the second of the above
options and provides (almost) vanishing values for both 
$\delta _{02}$ {\it and} $\delta _{13}$ through simple  and hence
 appealing mechanisms. Namely, in the $SL$ model
there are no  S - and   D -wave resonances, hence 
$\delta _{02} = 0$  and $\delta _{13} \propto PP'$. Moreover, 
in the $SL$ model the relation $\delta _{13} \simeq 0$ is
verified because the only 
$P_{11}$ resonance (Roper resonance) has a very tiny overall coupling 
in the process $\gamma p \to P_{11} \to K^+ \Lambda$; namely,  
 the product of the initial state 
(electromagnetic production vertex) and the final state 
(strong decay vertex) coupling constants  
$G_{N^*} \equiv g_{\gamma p N^{\star}} 
\cdot g_{K \Lambda N^{\star}}$, ($N^{\star} \equiv P_{11}$), comes out 
to be very small (see Ref.~\cite{SL} Tables~IX),  
as determined  by fitting the relevant data
(differential and total cross sections, 
the $\Lambda$-polarization asymmetry and the radiative capture branching ratio).
Our analysis hence  explains the negligible role
(see Ref.~\cite{SL} Tables~XII ) played by 
the Roper resonance in the strangeness  electromagnetic production reaction.

The  above discussion provides a clear example
of the  significant role that our nodal approach can play in establishing
links between data and dynamical models. More precisely, if
the  forthcoming polarization data, expected to be more accurate and 
contain more complete angular distributions, confirm the above analysis 
of the $n=3$ polynomial coefficients, then  future models could exclude the 
nucleonic $S_{11}$ and $P_{11}$ resonances from consideration,
 thereby decreasing considerably the number of 
candidate resonances, {\it and consequently}, the number of resonance sets 
to be investigated.

We now turn to the third case of pseudoscalar photoproduction
and introduce its resonance structure.

\subsubsection{ Eta}

New experimental facilities are or will soon be used to study
 $\eta$-photoproduction extensively.
 Recent low energy
cross section measurements~\cite{dyt95,kru95} have already provided  
insights into the dynamics of this process. At the present time, it
seems to be established~\cite{ben95}, {\it via} an effective Lagrangian 
approach including {\it s}-channel spin 1/2 \& 3/2 nucleonic resonances 
and {\it t}-channel vector meson exchange processes, 
that the reaction mechanism, at least for
$E_{\gamma}^{lab} \leq 800$, is dominated by two
resonances: $S_{11}(1535)$ and $D_{13}(1520).$
These data, as well as more extensive preliminary data from
ELSA~\cite{elsa,did96} between threshold and 1150 MeV,  have also
been investigated~\cite{aja96} within a formalism 
based on an isobar model~\cite{aja96,bou94}. 
In this approach electric and magnetic 
multipole amplitudes are expressed in terms of various isospin-1/2 
nucleonic resonances described by  ``relativized" energy-dependent 
Breit-Wigner 
forms,  plus a smooth background including S- and P- waves.
The role of the following resonances has been investigated:
$S11(1535)$, $S11(1650)$, $P11(1440)$, $P11(1710)$, $P13(1720)$, 
$D15(1675)$, $D13(1520)$, $D13(1700)$, $F15(1680)$, $G17(2190)$.
This isobaric approach of Ref.~\cite{aja96}, 
which is less fundamental than the effective Lagrangian formalisms,  
has the advantage of allowing one to rather easily 
include  higher spin resonances in the reaction mechanism.
The results of this isobar model 
work~\cite{aja96} confirm the major role played by the 
$S11(1535)$ and $D13(1520)$ resonance. More reliable conclusions
about the reaction mechanism up to 1.2 GeV  await the release
of final data.

Nevertheless, two main questions are worth investigating:
{\it i)} Can the sub-threshold, but wide, Roper resonance play a 
significant role in the reaction mechanism,
especially with respect to the forthcoming higher energy data~\cite{did96}
from Bonn?  {\it ii)} Could this process be used to search 
for undiscovered and/or missing resonances\footnote{
In this paper,  we focus on the following resonances predicted
by Capstick and Roberts\cite{cap94}: $P_{11},$  
$P_{13,}$  and $D_{15}$ with 
masses around 2 GeV, and non-vanishing decay amplitudes to the 
$\eta N$ channels.} 
as predicted~\cite{cap94} 
by recent relativized pair-creation quark models?

To address these questions,
 we single out the most relevant dynamical sets of resonances.
In Table~III, we list all observables and, using Table~II as input,
we indicate the resonance dependence of the polynomial expansion 
coefficients for various sets of assumed resonance amplitude scenarios. 
Since the $\eta$ photoproduction
is dominated by $S_{11}(1535)$ and $D_{13}(1520)$, we
start from the $SD$ resonant set, then we sequentially add in 
contributions from
 $P_{11}  \equiv P,$ 
$P_{13} \equiv P'   $ and 
$D_{15} \equiv D'.$  Thus, we consider the resonance scenarios of: 
 only $    SD,$ resonances,  then add in $P$ to get $    SPD$   
or $P'$ to get  
$  SP'D$, add in both spin 1/2 and 3/2 P-waves ($SPP'D$). 
 Finally, with $P,~P',~\&~D'$ all on, the resonance set is: 
 $ SPP'DD'.$  
This sequence of resonances has been generated by the 
following considerations: {\it i)} the main reasons for considering
 $P_{11}$ are that, first it is desirable to identify
 observables which could reveal the role,  if any, 
  played by the Roper resonance $P_{11}(1440)$ 
in this reaction, and secondly to look for two of the
missing $P$ resonances  with masses around 1.9 GeV; 
{\it ii)} for the $P_{13}$ sector, it is desirable to find out which
observables, if any, are suitable in searching  for 
 undiscovered resonances;  
{\it iii)} then a combination of these
four family of resonances 
({\it i.e.} $SPP'D$), and  an additional contribution from a spin 5/2 
resonance ($D'$) is investigated. 
 The need for such a 
spin-5/2 resonance ($D'$) has already been anticipated in
the case of pions~\cite{gar93}, and 
shown in the case of kaons~\cite{SL}.
 So, either a known or a missing $D'$ 
high spin resonance might also appear in the $\eta$ case in the comparable
energy region. If so, the resonance set $(SPP'DD')$ in Table~III
  should be considered
in determining the best observable for seeing the $D'$ effect in
$\eta$ photoproduction.

  For each of the  above $\eta-$nucleon resonance scenarios,
we  use Table~II to restrict the relative magnitudes of the $a_m$ 
coefficients for $\eta$ photoproduction.  
In  Table~III,  we summarize these $\eta$ photoproduction 
results.~\footnote{  
Similar specializations of Table~II could be generated
for the $\pi$ and $K^+$ cases.} 
The relations in Table III are generated from Table~II in the following 
manner: recall that  S is the dominant
resonance (85 to 90\% of the $\eta$ photoproduction cross-section) and 
then add in the other significant resonance 
D13. With just these two dominant resonances, we obtain vanishing values 
for some coefficients.  We also obtain the order of the polynomial that 
can be generated by just these two resonances--the $SD$ resonance set.
 Next, additional resonance scenarios are considered in the order
$SPD,$ $SP'D,$ $SPP'D,$ and finally $SPP'DD'.$  At each step we obtain the
possible order of the polynomial, along with some rules on the
polynomial coefficients $a_m.$  Most of the rules are based on the
dominance of $S$ and $D$ amplitudes,  followed by the secondary
P and D- resonances. In addition,  
   there is an additional  arbitrary assumption made at times that 
interference terms are all constructive. This assumption is 
of course not always true;  indeed,  we  saw  some
examples of destructive interference occur in the
pion and kaon reactions. For those special times, the (unequality) relations 
between polynomial coefficients in Table~III, provide
only ``upper" or ``lower" limits on the coefficients, assuming constructive 
(or slight destructive) interferences. However, even in the case of highly 
destructive interferences, as discussed in pion and kaon Sections, 
our approach allows extracting significant informations on the dynamics of 
the investigated processes.

As an example of how to obtain Table~III from Table~II,  consider the
$T$ term in Table~II. The coefficients for $T$ extracted from Table~II, 
under the assumption of zero $D',$  were presented earlier in Eq.~\ref{coefT}. 
For the $\eta$ case with just the $SD$ dominant resonances,  we see from 
Eq.~\ref{coefT}, that $a_0 \equiv 0,$ and that only $a_1$ is on, e.g. a 
polynomial of $n=1$ order is obtained.  
These facts are entered in Table~III in the $T$ row and the $n=1$ polynomial 
column.  Footnote $d$ presents the additional
information that $a_0 =0$ for this resonance scenario.  In this way,
all of Table~III  is generated.   
  
Information stored in Table~III for $\eta$ photoproduction
 relates not only to the angular structure of  observables,
but also, since higher $n's$ enter with increasing energy, to their
energy evolution.  We now wish to address the
question `Can a previously undetected resonance drive one of these
polynomial coefficients and dramatically alter the
angular and energy dependence of specific spin observables?'

Let us begin by examining the cross-section part of Table III. 
From their cross-section data,  Krusche {\it et al.}~\cite{kru95}
 concluded that only $S_{11}(1535)$ and $D_{13}(1520)$ resonances 
are required, which is also a feature of recent 
models~\cite{ben95,aja96}.
The absence of P-waves, especially  due to the Roper resonance,
has been deduced by  Krusche {\it et al.}~\cite{kru95} from their 
finding that a polynomial of the second order,  
with  $a_1 \approx 0$, (within the  experimental uncertainties),  
suffices to fit their cross-section data.
Both $SD$ and $SPD$
resonance scenarios lead to such a second order polynomial form for the
cross-section\footnote{The cross-section $\sigma,$  is a 
${\cal L}_0$ class observables; its profile function is
 ${\cal I}$ which, as indicated earlier, is defined by
$\sigma \equiv (q/p) {\cal I}.$},  however,  for the 
  case of  just the $SD$  resonance set one finds that $a_1=0$ from 
Table~III.
In contrast, a $SPD$ set of resonances yields  $a_0>a_2>a_1 ,$ which 
suggests that in a reaction mechanism {\it dominated} by  the $SD$ set, 
introduction of an additional secondary $P_{11}$ resonance 
(e.g., the $SPD$ set)  should  yield  a 
{\it small} $a_1$  coefficient,  compared to $a_0$ and $a_2 .$ 
Finding clean evidence for a $P_{11}$ resonance effect from  cross-section 
data, in a situation where $S$ and $D$ resonances dominate, thus requires 
one to extract an $a_1$ coefficient from the data with a {\it significant} 
non-vanishing value ({\it within the associated experimental errors}). 
That is a quite difficult experimental task and hence suggests that we go 
beyond the cross-section in searching for $P_{11}$ resonance effects.

There are observables that are more sensitive to $P_{11}$ resonance(s) 
effects than the cross section.
For example,  consider the single target ($T$) or 
recoil ($P$) polarization asymmetries, as well as the double polarization 
beam-target observables $H$ and $F $ (with linearly or circularly  
polarized beams, respectively).  All of these four spin observables share 
the property that both
$SD$ and $SPD$ sets lead to first order polynomials, 
$\sin^M\theta(a_0 + a_1 \cos\theta,$ see Table~III. In the case of the pure 
$SD$ resonance set,  we find that $a_0=0$;  hence, a node at $90^\circ$ is 
anticipated for all four of these observables. 
 With the $SPD$ resonance set, both coefficients of the first order polynomial
 are finite for all four of these observables,  with
$a_1>a_0, $ which  means that these observables have one node at 
$\theta_0 \not= 90^\circ.$ The deviation of the node position from 
$90^\circ$ depends on the ratio $a_0/a_1$ and therefore
are sensitive measures of  the importance of the $P_{11}$ amplitude.

 From among the four observables discussed above,  the  $P$ and 
$H$ asymmetries have a potentially useful property that all of the polynomial 
coefficients for $P$ and $H$  
arise {\it exclusively} from interference terms, see Table~II. Thus, the  $a_0$ 
coefficients for $P$ and $H$ observables  are   particularly excellent ways to 
amplify  P-wave  (both $P_{11}$ and $P_{13}$) effects,  since they appear 
interfering with the two dominant $S$ and $D$ resonances in $a_0.$

For the four observables $T, P, H\ {\rm and}\  F,$  investigation of the 
$SP'D$ set is also very informative.
Here, we are dealing with a second order polynomials with $a_1>>a_ 0>a_2.$ 
Hence,  besides the special sensitivity of  $a_0$ to P-waves, this
polynomial gets two roots and possibly two nodes.  Thus, observation
of double nodes in these observables,  especially if they evolve
rapidly with energy,  would be strong indication of a 
$P'$  resonance. 

In summary, for the observables $P$, $H$, $T$, and $F$, the resonance 
sets $SD$, $SPD$, $SP'D$  lead to one node at $90^\circ$, and one node 
at  $\theta \not= 90^\circ$, and  possibly two nodes, respectively. 
Direct experimental evidence of such nodes could be a way to reveal 
associated $P $ and/or $P'$ resonance dynamics.
 
For the $SPP'D$ case, we see from Table~III, that two nodes are possible 
for the observables  $P$, $H$, $T$, and $F.$  In these cases, however, 
$a_2 \not=0$ by itself (see Table~II) implies contributions from $P_{13}$ 
resonance(s), whether or not
$P_{11}$ resonances contribute.

The single photon polarization asymmetry, $\Sigma,$ and the 
double polarization beam-target observable, $G,  ( {\cal P}^{00}_{l z}) $ 
with linearly polarized beam,
show  no sensitivity to additional $P_{11}$ resonances,
 since both $SD$ and $SPD$
sets generate first order polynomials for these observables, with $a_1=0.$
 Hence for both $SD$ and $SPD$ resonance 
sets, $\Sigma,$ and   $G$  are nodeless.    
 Adding a $P_{13}$ to any of these sets ($SP'D$ and $SP'PD$) leads to
$a_0>a_1\not=0$, in which case $\Sigma$ and $G$ remain nodeless.
Hence, $\Sigma$ and $G$  are particularly insensitive, especially in their
nodal structure,  to $P'$
resonances.  They are, however, quite sensitive to the addition of a $D'$
resonance,  since it opens the possibility of two nodes.  A bifurcated nodal
trajectory in either  $\Sigma $ or $G$ ,  which involves going from zero 
to two nodes,  especially if it occurs rapidly,  could be striking evidence 
of a $D'$ effect.    
 
The beam-recoil asymmetries 
$O_{z'}  ( {\cal P}^{0z'}_{l 0})$ 
(with linearly polarized beam) and the target-recoil 
$T_{x'}  ( {\cal P}^{0x' }_{0 x} ) $ 
produce a first order polynomial for all resonance scenarios, except for 
the full case of $SPP'DD', $ wherein, the $D'$ resonance enters. 
For these observables a node at $90$ degrees occurs (since $a_0 \equiv 0 $) 
assuming just the pure $SD$ set.
The $SPD,$ $SP'D,$ and $SPP'D,$ scenarios all generate a one non-$90$ 
degree node situation. The full scenario set $SPP'DD',$ brings in a cubic 
polynomial,  which suggests that these observables could have bifurcating 
nodal trajectories, for which the change from one to three node is
driven by a $D'$ resonance.

For the  beam-recoil 
$C_{z'} ( {\cal P}^{0z'}_{c 0})$ 
(with a circularly polarized beam) and target-recoil 
$L_{z'} ( {\cal P}^{0z'}_{0 z} ) $ observables, we see from Table~III that     
 a polynomial of third order with $a_0=a_2=0$ is obtained in the case of a 
pure $SD$ resonance set.
Thus $C_{z'}$ and $L_{z'}$  have  $90$ degree nodes in that 
limit.\footnote{Both $C_{z'}$
and $L_{z'}$ must have an odd number of nodes according to the general
helicity rules of Ref.~\cite{FTS}. }  Since $a_1 > a_3$ is also
indicated for the pure $SD$ case,  a second node is unlikely.  However, 
with the addition of
$P$ and $P'$ resonances,  these observables could 
acquire up to three nodes for all non-$D'$ cases.  Once a $D'$ resonance 
enters, the polynomial jumps to fifth order,  with five nodes possible. 
 
Now consider the most complicated reaction mechanism presented in
Table III; namely, the case of the full resonance scenario $SPP'DD'.$
 As can seen in Table II, the highest
order coefficient for all observables is either a pure $D'$ state 
(${\bf C_{z'}}\ ({\cal P}^{0z'}_{c 0})$, 
${\bf L_{z'}}\ ({\cal P}^{0z'}_{0 z})$, 
${\bf C_{x'}}\ ({\cal P}^{0x'}_{c 0})$, 
${\bf L_{x'}}\ ({\cal P}^{0x'}_{0 z}) $,
${\bf O_{x'}}\ ({\cal P}^{0x'}_{l 0})$, 
${\bf T_{z'}}\ ({\cal P}^{0z'}_{0 x})$,
${\bf O_{z'}}\ ({\cal P}^{0z'}_{l 0})$,
${\bf T_{x'}}\ ({\cal P}^{0x'}_{0 x})$),   
or a combination of pure $D'$ state plus an amplification of it by
the $D$ resonance (  $D'\oplus DD'$  ) 
($d\sigma$, 
${\bf E}\ ({\cal P}^{00}_{cz})$,
${\bf T}\ ({\cal P}^{00}_{0y})$,
${\bf F}\ ({\cal P}^{00}_{c x})$,
${\bf \Sigma}\ ({\cal P}^{00}_{l0})$,
${\bf G}\ ({\cal P}^{00}_{lz})$).
or just a $ DD'$ interference term--which is the case for   
${\bf P}\ ({\cal P}^{0y'}_{00})$, 
${\bf H}\ ({\cal P}^{00}_{ l x})$, 
These last eight observables,
by virtue of their $DD'$ terms, allow the dominant $D$
to overlap and hence magnify the
role of a possible $D'$ resonance. Thus they offer particularly suitable 
observables for investigating the contributions of any known or missing 
$D' \equiv D_{15}$ resonances.

Finally, we emphasize that, since the relations among the polynomial 
coefficients displayed in Table~III, often give clear information 
about the anticipated nodal and polynomial structures
 of the relevant observables, 
there is hope that the character of angular distributions and associated
nodal structure might yield definitive evidence for specific resonances.


\section{Summary and Conclusions}
 
Our analysis of the angular distribution of forthcoming 
polarization observable data and their 
special nodal trajectory and polynomial characteristics, within a 
multipole truncated basis, offers a potentially powerful means for  
investigating the underlying dynamics of  pseudoscalar
meson  ($\pi$, $K$, $\eta$) photoproduction. This method provides 
a helpful guide for  
phenomenological dynamical approaches by singling out the appropriate
families of nucleonic resonances
required by existing data.  Also, confronting this polynomial
expansion analysis with existing phenomenological approaches,  
allows one to emphasize both the strong and 
weak points of such models and put forward suggestions
for improvements. Moreover, this method promises to be a
  helpful guide in  planning experiments to search for missing and/or
yet undiscovered resonances,  
which constitutes 
a crucial test of quark-based descriptions~\cite{cap94,cap93,kon80}
of baryon spectra.

While waiting for a new generation of precision data,   we confronted our
 approach with both extant, but scarce,  polarization 
data,  and  with the predictions obtained using recent phenomenological
models.  In the process,  we learn  some things.  
For example,  for pion photoproduction, 
 we focused  on the polarized target asymmetry $T$ data~\cite{BONN}, 
 and showed that our approach incorporates some of the established facts;
namely,  that the reaction is dominated by the $\Delta_{33}$
resonance with non-negligible contributions from other spin 1/2
and 3/2 resonances. We also ascertained that the 
single beam polarization asymmetry ${\bf \Sigma}$
(${\cal P}^{00}_{l 0}$), and the double beam-target observable 
${\bf G}$ (${\cal P}^{00}_{l z}$) (both requiring a linearly 
polarized beam) are the best observables for investigating the 
suspected contributions of spin 5/2 resonances~\cite{gar93}.

We also examined the strangeness photoproduction reaction,  which has 
a rather complicated reaction mechanism~\cite{AS,WJC,SL,li}. 
We focused on the only available hyperon recoil polarization 
asymmetry $P$ (${\cal P}^{0y'}_{0 0}$) data~\cite{ELSA}, 
and  showed that, if these data are confirmed by future  
experimental results, then {\it s}-channel
spin 1/2 \& 3/2 resonances are  not playing a role in the reaction. 
This possibility would then
  considerably simplify the number of  
resonances needed in kaon photoproduction. 

New physics comes in while investigating the $\eta$ meson photoproduction
process. Recent low-energy measurements~\cite{dyt95,kru95} and the
preliminary higher energy data~\cite{elsa} of $\eta$ photoproduction
 lead to a very simple reaction
mechanism~\cite{kru95,ben95,aja96}; namely,  the reaction
seems to be dominated by two nucleonic resonances the $S_{11}(1535)$ and 
$D_{13}(1520).$ 
This dominance suggests using $\eta$ meson photoproduction to 
search for at least a few of  the missing and/or undiscovered resonances, 
which have been predicted~\cite{cap94} to couple to the $\eta N$ rather than
to the $\pi N$ channels.  This reaction offers a test of $QCD$ inspired 
models~\cite{cap94,kon80}; namely, of predicted $P_{13}$ 
and to a lesser extent $D_{15}$ resonances  
with masses below,
or slightly above, $2$ GeV. For the $P_{13}$ resonance(s) the 
recoil polarization $P,$ which is probably  rather difficult to  
measure, and the double polarization observable 
$H$ (${\cal P}^{00}_{l x}$)
(which requires   a polarized target
and a linearly polarized photons beam) are highly appealing.
These observables offer a similar and unique selectivity among
the sixteen observables; namely, their multipole polynomial expansion
coefficients depend {\it only} on interference terms. Hence,
the contributions from a sought-after $P_{13}$ resonance is magnified by 
the two dominant amplitudes. Moreover, the presence of a $a_2\not=0$ term
in the second or third order polynomials serve as  an unambiguous signature
for the presence of, at least, one $P_{13}$ resonance; 
with two nodes expected in the case
of a second order polynomial. This reasoning applies also to two other 
observables; namely,
the single target asymmetry $T$ and the  double polarization observable
$F$ (${\cal P}^{00}_{c x}$) 
(which involves  polarized target and a circularly polarized photon beam.)

Another interesting problem in the $\eta$ case concerns the
role played, if any, by the Roper-resonance $P_{11}(1440).$  
Our approach shows that the differential cross-section is not
sensitive enough to the Roper-resonance. However, four polarization 
observables are very suitable for this purpose. They are, as  
above, the single target asymmetry $T,$ 
 the recoil polarization $P,$  and the double beam-target polarization 
observables $H  \&  F.$ 
In contrast to the $P_{13}$ contributions leading to two nodes,
the Roper-resonance will produce only one node 
   and the deviation of that node away from
$90^\circ$ will give a measure of  
  the importance of the $P_{11}$ amplitude relative to
 the  $SD$ dominant resonances.

Finally, the effect of the spin 5/2 $D$-resonances, known or missing,
show up clearly in the highest order coefficients 
for any of the three single polarization observables,
or for any beam-target double polarization asymmetry($E$, $F$, $G$, $H$). 

In summary,  some puzzles in  hadron spectroscopy might
be answered by studying  $\eta$ photoproduction. 
Our investigation shows that the most promising observables require asymmetry
measurements with polarized beam and/or polarized target. So, final results 
from the recent polarized target asymmetry $T$ measurements~\cite{ant96} 
at ELSA are awaited for. Moreover, 
polarized beams are becoming available at CEBAF and GRAAL and new 
advances in the polarized target techniques
~\cite{did94} are expected to render such single and even double 
polarization measurements feasible in the near future.  

\acknowledgments
We wish to thank J. Ajaka, C. Fayard, P. Girard, P. Hoffman-Rothe,
and G.H. Lamot for helpful exchanges, and J.P. Didelez for his
interest in this work and much encouragement. 
Each of us wishes to express appreciation for 
warm hospitality  during visits to
the University of Pittsburgh (B.S.) and to Saclay (F.T.).
\begin{appendix}
\section{Multipole Expansion for the Target Polarization}
The polarized target asymmetry profile function is given here in terms of 
the polynomial expansion.  The associated
coefficients $a_m$ are then expressed as imaginary  parts
of bilinear products of multipole amplitudes.  This case is
used to illustrate the compact notation used in Table~I,  
 wherein  the general structure of
spin observables for pseudoscalar meson photoproduction is displayed.
The profile function $T(\theta)$ is of Legendre class ${\cal L}_{1b}$
and hence has the general form:
\begin{equation}
T(\theta) \equiv {\cal O}^{00}_{0y} (x) = \sin \theta \, 
\sum_{m=0}^n a_{m}x^m ,\nonumber \\ [7pt]
\end{equation} 
 with 
$x \equiv\cos(\theta).$  The polynomial expansion coefficients
are expressed in terms of electric and magnetic multipole
amplitudes as:
\begin{eqnarray} 
a_0\    &  =  {\rm Im}  &         \,  [ \,    E^{0}_{+}
[ - 3\, E^{+}_{1} + 3\, M^{+}_{1} ]^{*}- M^{-}_{1}
[ 3\,  E^{-}_{2} + 3\, M^{-}_{2} + 3\, E^{+}_{2} - 3\, M^{+}_{2} ]^{*}
\nonumber \\ [7pt] &  & 
+ E^{+}_{1}
[ - 6\,  E^{-}_{2} - {{27\,}\over 2} \, E^{+}_{2} ]^{*}
+ M^{+}_{1}
[ - 6\, M^{-}_{2} + {{15\,}\over 2} \, E^{+}_{2} + 6\, M^{+}_{2} ]^{*} 
\, \,  ]  \nonumber \\ [7pt]  
a_1\    &  = {\rm Im}  &      {{3\,}\over 2}   \,  [ \,  E^{0}_{+}
[  2\, E^{-}_{2} + 2\, M^{-}_{2} - 8\, E^{+}_{2} + 8\, M^{+}_{2} ]^{*}
- M^{-}_{1}
[ - 2\, E^{+}_{1} + 2\, M^{+}_{1} ]^{*} 
\nonumber \\ [7pt] &  & 
+ E^{-}_{2}
[ 25\, E^{+}_{2} + 2\, M^{+}_{2} ]^{*}
+ M^{-}_{2}
[ - 3\, E^{+}_{2} + 30\, M^{+}_{2} ]^{*} 
\nonumber \\ [7pt] &  & 
+ 8\, E^{+}_{1} \, M^{+*}_{1}
+ 8\, E^{-}_{2} M^{-*}_{2} \, - 18\, E^{+}_{2} M^{+*}_{2} 
\, \,  ]  \nonumber \\ [7pt]  
a_2\    &  = {\rm Im} &     {{3\,}\over 2}    \,  [ \,  M^{-}_{1}
[ 10\, E^{+}_{2} - 10\, M^{+}_{2} ]^{*}
+ E^{+}_{1}
[  12\,  E^{-}_{2} - 3\, E^{+}_{2} + 30\, M^{+}_{2} ]^{*}
\nonumber \\ [7pt] &  & 
+ M^{+}_{1}
[ 12\, M^{-}_{2} - 25\, E^{+}_{2} - 2\, M^{+}_{2} ]^{*} 
\, \,  ]  \nonumber \\ [7pt]  
a_3\    &  ={\rm Im} & {{45\,}\over 2} \, [ \, - 3\, E^{-}_{2} \, E^{+*}_{2} 
+ M^{-}_{2}
[ E^{+}_{2} - 4\, M^{+}_{2} ]^{*} 
 + 6\, E^{+}_{2} M^{+*}_{2} 
\, \,  ]  \nonumber \\ [7pt]  
\end{eqnarray} Expressions for all other observables are available
in reference~\cite{GST}.
\end{appendix}

\newpage
%
\begin{table}
\caption{The notation  
$ {\cal P}^{meson,\ final\ baryon}_{photon,\ initial\ baryon} $ is used
to indicate the initial (final) baryon  $ x y z(x' y' z')$ spin directions
and the photon's circular($c$) or linear($l$) polarization.
}
\begin{tabular}{lllcll}
Class&Observable & Symbol& & Notation & \\
\tableline    
{${\cal L}_0$ }& & & & & \\
\tableline
& {\bf Cross-section} & ${\bf I}$ & & ${\cal P}^{00}_{0 0}$ & \\  
& {\bf Beam-Target} & ${\bf E}$ & & ${\cal P}^{00}_{c z}$ & \\
& {\bf Beam-Recoil} & ${\bf C_{z'}}$ & &  ${\cal P}^{0z'}_{c 0}$ & \\ 
& {\bf Target-Recoil} & ${\bf L_{z'}}$ & & $ {\cal P}^{0z'}_{0 z}$ & \\ \\
\tableline 
{${\cal L}_{1a}$}& & & & & \\
\tableline
%
& {\bf Recoil} &  ${\bf P}$ & &  ${\cal P}^{0y'}_{00}$ & \\ 
& {\bf Beam-Target} & ${\bf H}$ & &  ${\cal P}^{00}_{ l x}$ & \\
%
& {\bf Beam-Recoil} & ${\bf C_{x'}}$ & &  ${\cal P}^{0x'}_{c 0}$ & \\ 
& {\bf Target-Recoil} & ${\bf L_{x'}}$ & &  ${\cal P}^{0x'}_{0 z} $ & 
\\ \\
\tableline 
{${\cal L}_{1b}$}& & & &  & \\
\tableline
%
& {\bf Target} &  ${\bf T}$ & &  ${\cal P}^{00}_{0y}$ & \\
& {\bf Beam-Target} & ${\bf F}$ & &  ${\cal P}^{00}_{c x}$ & \\
%
& {\bf Beam-Recoil} & ${\bf O_{x'}}$ & &  ${\cal P}^{0x'}_{l 0}$ & \\
& {\bf Target-Recoil} & ${\bf T_{z'}}$ & &  ${\cal P}^{0z'}_{0 x}$ & 
\\ \\
\tableline 
{${\cal L}_{2}$}& & & & &  \\
\tableline
%
& {\bf Beam} & ${\bf \Sigma}$ & &  ${\cal P}^{00}_{ l 0}$ & \\
& {\bf Beam-Target} & ${\bf G}$ & &  ${\cal P}^{00}_{l z}$ & \\
%
& {\bf Beam-Recoil} & ${\bf O_{z'}}$ & &  ${\cal P}^{0z'}_{l 0}$ & \\
& {\bf Target-Recoil} & ${\bf T_{x'}}$ & &  ${\cal P}^{0 x'}_{0 x}$ & 
\\ \\
\end{tabular}
\end{table}
%
%
\widetext 
\begin{table}[t]
\caption{Multipole dependence of the polynomial
coefficients for all single and double polarization
observables in pseudoscalar meson production. Here $ i=1$
and $i=3$ denote the $J=1/2$ and $3/2$ P-waves ($P \equiv P_{2I   1}$ 
and $P'  \equiv P_{2I   3},$ 
respectively);
while $j=3$ and $j=5$ are the $J=3/2$ and $5/2$ D-waves 
($D \equiv D_{2I   3}$ and $D' \equiv D_{2I   5}$, respectively).
 Single letters refer to terms of the type $|E^+_0|^2 \equiv S;$
such terms are listed in the first row for each set of observables,
when appropriate.
 In the following rows the interference terms
are given in the notation $S D_{2I j}, \cdots.$ 
The term $P_{2I   i} D_{2I   j}$ is short for
$ PD\oplus PD'\oplus P'D\oplus P'D'.$  The boxed letters
 show how the strong S-wave contributes to the $a_m$ coefficients for
each observable. }
\begin{tabular}{lllllllll}
 &$a_0$ & $a_1$ & $a_2$ & $a_3$  & $a_4$ & $a_5$ \\
\tableline    
{${\cal L}_0$ } & & & & &  &\\
\tableline
$d\sigma$ \& ${\bf E}$&
$\fbox{S}\oplus P_{2I   i}\oplus D_{2I    j}\oplus$ &
 & $P'\oplus D_{2I   j}\oplus$  &  & $D' \oplus$ &  \\
 & $ \fbox{$S D_{2I   j}$}\oplus P P' \oplus$&
$ \fbox{$S P_{2I   i}$}\oplus P_{2I   i} D_{2I   j}$ &
$ \fbox{$S D_{2I   j}$}\oplus PP'\oplus$ & $ P D'\oplus P' D_{2I   j}$& 
  $ DD' $&\\ 
& $ D D' $&  & $ DD'$ &  &   &\\ \\
${\bf C_{z'}} $ \& ${\bf L_{z'}}\ $& &
$\fbox{S}\oplus P_{2I   i}\oplus D_{2I   j}\oplus$ & &
$ P'\oplus D_{2I   j}\oplus$& & $D'$ \\
& $\fbox{$S P_{2I   i}$}\oplus P_{2I   i} D_{2I   j}$ &
$ \fbox{$S D_{2I   j}$}\oplus P P'\oplus$ &
$ \fbox{$SP'$}\oplus P_{2I   i}D_{2I   j}$ &
$ \fbox{$S D'$}\oplus  DD'$ &  $P'D'$  & \\
& & $ D D'$ & & &  & \\ 
\tableline 
{${\cal L}_{1a}$} & & & &  & &\\
\tableline
%
 ${\bf P}$ \& ${\bf H}$ & & & &  & &\\
& $\fbox{$S P_{2I   i}$}\oplus P_{2I   i} D_{2I   j}$ &
$\fbox{$S D_{2I   j}$}\oplus P P'\oplus$ & $P D'\oplus P' D_{2Ij} $&
$D D'$&  & & \\
 & & $  D D'$ & & &  & & \\ \\
%
${\bf C_{x'}}$ \& ${\bf L_{x'}} $&
$\fbox{S}\oplus P_{2I   i}\oplus D_{2I   j}\oplus$ & &
$P'\oplus D_{2I   j} \oplus$ & & $D'$  & \\
&  $\fbox{$S D_{2I   j}$} \oplus P P' \oplus$&
 $ \fbox{$S P'$}\oplus P D_{2I   j}\oplus $ &
$ \fbox{$S D'$}\oplus DD'$ & $ P' D'$ &   &  \\
&$DD'$ & $ P' D'$ & & &   &  \\ 
\tableline 
{${\cal L}_{1b}$} & & & &  & & &\\
\tableline
%
 ${\bf T}$ \& ${\bf F}$ & & $P'\oplus D_{2I   j} \oplus$&
 & $D'\oplus$&  & &\\
& $\fbox{$S P'$}\oplus P_{2I   i} D_{2Ij}$ &
$ \fbox{$S D_{2I   j}$}\oplus P P'\oplus$ &
$ P D'\oplus P' D_{2I   j}$ & $ D D'$ &    &  &  \\
& & $ D D'$ & & &   &  &  \\ \\
%
${\bf O_{x'}}$ \& ${\bf T_{z'}}$&
 $P'\oplus  D_{2I   j}\oplus$& &
$P'\oplus  D_{2I   j}\oplus$ && $D'$  & &\\
& $ \fbox{$S D_{2I   j}$}\oplus P P'\oplus $ &
$ \fbox{$S P'$}\oplus P_{2I   i} D_{2I   j}$ &
$ \fbox{$S D'$}\oplus D D'$ & $  P' D'$ & & & \\
& $ D D'$  &  & & & & & \\ 
\tableline 
{${\cal L}_{2}$} & & & &  & & &\\
\tableline
%
  ${\bf\Sigma}$ \&  ${\bf G}$&
$P'\oplus D_{2I   j}\oplus$ & & $D'\oplus$ & & & &\\
&$ \fbox{$S D_{2I   j}$}\oplus P P'\oplus$ &
$ P D'\oplus P' D_{2I   j} $ &$ D D'$ & & & &\\
& $ D D'$ & & & & & &\\ \\
%
${\bf O_{z'}}$ \& ${\bf T_{x'}}$&
 &$P'\oplus D_{2I   j}\oplus$ & & $D'$&  & \\
& $ \fbox{$S P'$}\oplus P_{2I   i} D_{2I   j}$ &
$ \fbox{$S D'$}\oplus D D' $ & $ P' D'$ & &  & &  
\end{tabular}
\end{table}
 
%
\begin{table}
\caption{The role of various resonance scenarios on  $\eta$ photoproduction.
 The amplitudes are assumed to be dominated by resonances. 
Starting from the well-known resonances
$SD(S=S_{11}(1535)$ and $D_{13}(1520)$), others are added sequentially;
 namely, $P\equiv P_{11}(1440),$ 
$P'\equiv P_{13},$ $D\equiv D_{13},$ $D'\equiv D_{15} $ to
generate various resonance scenarios.  For each combination  of resonances, 
and for polynomial orders of $n=1, (a_0 + a_1 \cos\theta); n=2 , 
(a_0 + a_1 \cos\theta + a_2 \cos^2\theta),$ etc.; 
the relative size of the expansion coefficients are predicted, based on
Table~I. 
A large coefficient is denoted as $a_m>>.$  This information can be used to 
predict the effect of a given set of resonances on the angular and energy 
dependence of spin observables for $\eta$ meson photoproduction.  
}
\begin{tabular}{llllllll}
 Observable& $n=1$ & $n=2$ & $n = 3$ & $n = 4$ & $n = 5$\\
\tableline    
{${\cal L}_0$ } &  &  &  &&\\
\tableline
$d\sigma$ \& ${\bf E}\ $
& & $SD^a $ & $SP'D^g $ & $SPP'DD'$& \\
& &$SPD^b$ & $SPP'D^g $ & & \\ \\
${\bf C_{z'}}$ \& 
${\bf L_{z'}}$
& & & $SD^c$ & &$SPP'DD'$ \\  
&  & &$SPD$  & & \\
&  & & $SP'D$ & & \\
&  & & $SPP'D$ & & \\ 
\tableline 
{${\cal L}_{1a}$} & & &   & & \\
\tableline
%
 ${\bf P}$ \& 
${\bf H}$
&$SD^d$ & $SP'D^e$ & $SPP'DD' $  & & \\
&$SPD^f$ &$SPP'D^e$ &  & & \\ \\

%
${\bf C_{x'}} $ \& 
${\bf L_{x'}} $
& &$SD^a$   & 
&$SPP'DD'$   & \\
& &$SPD^b$ &  & & \\
&  & $SP'D^b $ &  & & \\
&  & $SPP'D^b $ &  & & \\ 
\tableline 
{${\cal L}_{1b}$} & & &   & & \\
\tableline
%
 ${\bf T}$ \&
${\bf F}$
& $SD^d$  & $SP'D^e$ & $SPP'DD'$  & & \\ 
& $SPD^f$  & $SPP'D^e $ &  & & \\ \\
%
${\bf O_{x'}}$ \&
${\bf T_{z'}}$
& &$SD^a$ & $SP'D^b$  & $SPP'DD' $& \\
& & $SPD^b$ & $SPP'D^b$ & & \\  
\tableline 
{${\cal L}_{2}$} & & &   & & \\
\tableline
%
  ${\bf \Sigma}$ \& ${\bf G}$
& $SD^h$ & $SPP'DD'$ &  & & \\
& $SPD^h$ & &  & & \\
& $SP'D^i$ & &  & & \\
& $SPP'D^i$ & &  & & \\ \\
%
${\bf O_{z'}}$  \&
${\bf T_{x'}}$
& $SD^d $ & & $SPP'DD' $  & & \\
& $SPD^j $ & &   & & \\
& $SP'D^j$ & &  & & \\
& $SPP'D^j$ & &  & & \\ 
\end{tabular}
$^aa_0>a_2,~a_1=0$; $^ba_0>a_2>a_1$; $^ca_0=a_2=0,~a_1>a_3,~a_3\,
{\rm pure}\ D-{\rm wave}$
; $^da_0=0$; $^ea_1>>a_0>a_2$; $^fa_1>a_0$; $^ga_3 \propto P'D$; 
$^ha_1=0$; $^ia_0>a_1$, $a_1 \propto P'D$; $^ja_1>a_0$

\end{table}

%

\begin{figure}

\caption{Typical energy and angular dependence of a 
${\cal L}_0$ class polarization
observable asymmetry (the depicted case is for a typical
 double polarization observable $E$ for kaon production).
 The nodal trajectory is defined as the
projection of non endpoints zero values at each energy on the plane 
defined by the energy of the incident photon and the angle of
produced meson.
}
\end{figure}

\begin{figure}

\caption{Angular distribution of the polarized target asymmetry 
in the $\gamma \vec{p} \rightarrow  \pi^+ n$ reaction at
$E_{\gamma}^{\rm lab}=220$ MeV ($a$) and $650$ MeV ($b$).
Curves are explained in the text.}
\end{figure}
\begin{figure}
\caption{Angular distribution of the recoil $\Lambda$-polarization 
asymmetry in the  
$\gamma p \rightarrow K^+ \vec{\Lambda}$ channel at 
$E_{\gamma}^{\rm lab}=1.2$ GeV. Curves are explained in
the text.}
\end{figure}
\end{document}